\documentstyle[12pt, aasms4, graphics]{article}

\begin{document}

\title{EXPLAINING THE GAMMA RAY BURST LAG/LUMINOSITY AND
VARIABILITY/LUMINOSITY RELATIONS}

\author{Bradley E. Schaefer\\
University of Texas, Department of Astronomy, C-1400, Austin TX
78712}

\baselineskip 12pt

\begin{abstract}

Two different luminosity indicators have been recently proposed for Gamma
Ray Bursts that use the lag between light curve peaks with hard and soft
photons as well as the variability ('spikiness') of the burst light
curve.  Schaefer, Deng, \& Band have proven that both of these luminosity
indicators are valid by finding a lag/variability relation for 112 BATSE
bursts.  Here, I provide simple and general explanations for both
indicators.  The lag/luminosity relation is a consequence of conservation
of energy when radiative cooling dominates.  High luminosity bursts cool
rapidly so the lag between when hard and soft photons peak is short, while
low luminosity bursts cool slowly so the lag is long.  The
variability/luminosity relation arises within internal shock models
because the luminosity and variability both scale as strong functions of
the bulk relativistic motion of the jet, $\Gamma$.  With luminosity
roughly
proportional to $\Gamma^5$ and variability scaling as $\Gamma^2$, we get
the luminosity
being correlated to variability just as the observed power law.

\end{abstract}


\clearpage

\section{INTRODUCTION}

One of the most fundamental problems of Gamma-Ray Burst (GRB) studies is
to establish the distance scale.  The recent discoveries of x-ray,
optical, and radio transients have allowed only a dozen or so red shifts
to be measured.  At the Fifth Huntsville Gamma-Ray Burst Symposium, two
luminosity (and hence distance) indicators were proposed that make use of
gamma ray light curves only and hence can be used for large numbers of
long-duration bursts (Norris, Marani, \& Bonnell 1999; Fenimore \&
Ramirez-Ruiz 2000).  The first indicator is  $\tau_{lag}$, which is the
time delay
between peaks in the light curve as viewed with hard and soft
photons.  More specifically, the lag is the delay of the maximum
cross-correlation between the burst light curves in BATSE channels 1
(25-50 keV) and 3 (100-300 keV).  The second indicator is the variability,
$V$, which is the 'spikiness' of its light curve, taken as the normalized
variance of the light curve around a smoothed light curve of the
burst.  Both indicators were calibrated by only 6 or 7 bursts with known
red shifts; however, Schaefer, Deng, \& Band (2001) have proven that
{\em both}
the lag/luminosity and the variability/luminosity relation are correct by
finding the required lag/variability relation with 112 BATSE bursts.

Before the two luminosity indicators can be used with confidence, we
should have some physical understanding of {\em why} they both work.  This
{\it Letter} gives a simple explanation for both relations with standard
assumptions.

\section{BACKGROUND}

I will presume that GRBs are internal shocks produced by narrow
jets expanding with a relativistic $\Gamma$ (roughly 100-1000) with a
total mass
$M_{jet}$ (with a kinetic energy that does not vary greatly between
bursters) with an opening angle $\Theta_{jet}$ (such that
$\Theta_{jet} < \Gamma^{-1}$) for which the
cooling is dominated by radiation.

The luminosity is calculated as 
\begin{equation}
        L = 4 \pi D_L^2 \cdot P_{256} \cdot <E>,
\end{equation}
where $P_{256}$ is BATSE's observed peak flux (from 50-300 keV evaluated
on the
256 ms time scale), $D_L$ is the burst's luminosity distance 
(for $H_o = 65 km \cdot s^{-1} \cdot Mpc^{-1}$, $\Omega_M = 0.3$, and
$\Omega_{\Lambda} = 0.7$),
and $<E>$ is the average energy of a photon
for an $E^{-2}$ spectrum ($1.72 \times 10^{-7} erg \cdot
photon^{-1}$).  This formula includes a
K-correction for an $E^{-2}$ spectrum as appropriate for average bursts
(Schaefer et al. 1994, 1998).  These luminosities implicitly represent the
case when the radiation is emitted isotropically.

The existence of the fairly tight lag/luminosity relation (and the less
tight variability/luminosity relation) suddenly makes GRBs into standard
candles.  This is not in the sense that all bursts have the same
luminosity, but in the sense that they have a light curve property that
can be used to uniquely determine their luminosity.  In a similar sense,
Cepheids are standard candles not because they all have the same
luminosity but since the period/luminosity relation allows their
luminosity to be known from a light curve property.  Similarly, Type Ia
supernovae are standard candles with their luminosity determined by their
decline rate.

With GRBs as standard candles, there must be something 'standard' imposed
by the physics of the problem.  Within the general picture of internal
shocks, the luminosity will scale something like the energy of the jet
($\Gamma M_{jet} c^2$) perhaps with additional factors of
$\Gamma$.  This is supported by
Kulkarni (2000) who claims that the total energy budget of bursts is
constant.  It might not be surprising to have a constant quantity between
GRBs, since bursts may be a critical phenomenon such that they are
possible only when this quantity is above some threshold value and the
frequency of that value falls off sharply above that threshold hence
producing a near constant value.  In any case, some such quantity must be
roughly a constant from burst-to-burst for the lag/luminosity relation to
be true.

Three angular scales are important for the internal shock scenario.  First
is the relativistic Doppler beaming where the radiation is sent into a
forward cone with a characteristic opening angle of $\Gamma^{-1}$.  At
least a
substantial fraction of bursts must have $\Gamma$ greater than $\sim 100$
since
otherwise it will be difficult to get prompt GeV photons out of the shock
region.  Second is the jet opening angle, $\Theta_{jet}$, which might be
perhaps 10 degrees
or much smaller than 1 degree.  Third is the characteristic angular size
of the
disconnected emitting regions as viewed from the central source,
$\Theta_{region}$.  This angle must be very small since the majority of
bursts have
millisecond variability (Walker, Schaefer, \& Fenimore 2000), the average
light curve during decline does not show curvature effects (Fenimore
1999), and three bursts with good blackbody spectra have very small
blackbody radii for their emitting regions.  Quantitative analysis of
these three constraints show that $\Theta_{region}$ is as small as
arc-minutes.

At a great distance from the burster, the beam pattern will be a
convolution of the jet opening angle and the relativistic beaming.  The
characteristic angle of this will be roughly the larger of $\Theta_{jet}$ 
and $\Gamma^{-1}$,
for a total solid angle of $\Omega$.  I would expect that $\Theta_{jet}$
is not greatly
larger than $\Theta_{region}$, so then $\Theta_{jet} < \Gamma^{-1}$ and
$\Omega = \pi \Gamma^{-2}$.  In this case, an
observer on Earth will be able to see the entire working surface of the
jet and all the emitting regions, with none being hidden by relativistic
beaming.  Unless some arrangement of conspiring factors is invoked, the
visibility of the entire jet is a requirement for the existence of any
tight luminosity relation (such as the lag/luminosity relation).  That is,
otherwise the fraction of the available jet energy detected by BATSE will
vary with $\Gamma$ from burst-to-burst creating a large scatter in the
lag/luminosity relation.

\section{LAG/LUMINOSITY RELATION}

The lag/luminosity relation (Norris, Marani, \& Bonnell 1999) is
\begin{equation}
        L = 2.9 \times 10^{51} (\tau_{lag}/0.1s)^{-1.14 \pm 0.20},
\end{equation}
where $L$ is the observed peak luminosity in units of $erg \cdot s^{-1}$
(Schaefer,
Deng, \& Band 2001).  

The average cooling rate per particle in the emitting region of the jet
will be the total luminosity over all directions ($L_{tot}$) of the jet
divided
by the number of emitting particles.  The number of emitting particles
will be roughly $M_{jet}/m_{proton}$.  The cooling rate will also be the
time
derivative of the mean particle energy, $E_{peak}$.  So
\begin{equation}
	dE_{peak}/dt = -L_{tot}/(M_{jet}/m_{proton}),
\end{equation}
which is really just a statement of the conservation of energy for when
radiative cooling dominates.  Equation 3 is valid both in the reference
frame of the jet as well as the reference frame of BATSE (Liang
1997).  The proportionality between $dE_{peak}/dt$  and $L$ has previously
been
discovered from BATSE data by Liang \& Kargatis (1996) as a general law
covering the decay of many GRB pulses.

Let us evaluate equation 3 at the time of peak flux.  On the
right-hand-side, 
\begin{equation}
	L_{tot} = L \cdot (\Omega /4 \pi),
\end{equation}
where $\Omega$ is the solid angle into which the radiation is beamed
(roughly
$\pi \Gamma ^{-2}$ since $\Theta_{jet} < \Gamma^{-1}$).  On the
left-hand-side of equation 3, we can
approximate the derivative as a finite difference evaluated between times
when the light curves peak in BATSE channel 3 and channel 1;
\begin{equation}  
	dE_{peak}/dt = [E_{peak}(T_1)-E_{peak}(T_3)]/[T_1-T_3].
\end{equation}
The time when the light curve peaks in BATSE channel 3, $T_3$, is a
competition between the turn on of the pulse versus the cooling of the
emitting region and will be approximately when $E_{peak}$ is centered in
channel 3 at $\sim 200 keV$.  Similarly, $E_{peak}(T_1)$ will be near the
center
energy of the BATSE channel 1 ($\sim 30 keV$) at the time when the flux in
this
channel reaches maximum.  The time difference between the peaks in
channels 1 and 3 ($T_1-T_3$) is equal to $\tau_{lag}$, so
\begin{equation}
	dE_{peak}/dt = -170 keV/ \tau_{lag}.
\end{equation}
With eqs 3, 4, and 6, we get
\begin{equation}
	L = (4 \pi M_{jet} \cdot 170 keV \cdot m_{proton}^{-1}
        \Omega^{-1}) \tau_{lag}^{-1}.
\end{equation}
Thus, we find that the observed luminosity should be inversely
proportional to $\tau_{lag}$.  To within the fairly small uncertainties in
the
exponent of eq. 2, this reproduces the empirical lag/luminosity relation.

The existence of the lag/luminosity relation implies that some quantity
must be fairly constant from burst-to-burst, and equation 7 shows that
this quantity is $M_{jet}/\Omega$.  For $\Theta_{jet} < \Gamma^{-1}$,
$\Omega$ will be $\pi \Gamma^{-2}$ and so the nearly
constant quantity scales as $\Gamma^{2}M_{jet}$.

A comparison of equations 2 and 7 requires that the constants are
approximately equal.  This gives a value for the emitting mass in the jet;
\begin{equation}
	M_{jet} = 0.89 M_{\odot} (\Omega /4 \pi).
\end{equation}
For $\Gamma = 100$ (and hence $\Omega \sim 3 \times 10^{-4}$), $M_{jet}$ 
is $2.2 \times 10^{-5} M_{\odot}$.  This is just 7 Earth
masses.

To recap, my explanation of the lag/luminosity relation is that
the rate of cooling of the emitting region (and hence the lag) will depend
on the rate at which the energy is radiated away (and hence the
luminosity).  When the burst has a high luminosity, then the individual
particles in the emitting region are radiating their energy rapidly and
cooling quickly so the time from when $E_{peak} \sim 200 keV$ until the
time when $E_{peak} \sim 30 keV$ (i.e., the lag time) is short.  When the
burst has a low
luminosity, the particles are cooling slowly and the lag time to cool from
200 keV to 30 keV is long.

GRB980425 (associated with SN1998bw [Galama {\it et al.} 1999]) has a
measured
peak luminosity ($2 \times 10^{46} erg \cdot s^{-1}$) which falls roughly
five orders of
magnitude below the luminosity of typical bursts.  The observed very high
energy photons from ordinary GRBs implies that $\Gamma \sim 100$ or
somewhat higher,
so with $L \propto \Gamma^5$ (see below) the $\Gamma$ for GRB980425 must
be ~10 or somewhat
higher.  The lag of GRB980425 is 4 seconds, whereas its luminosity implies
a lag of 3400 seconds from equation 2.  This suggests that the emitting
region cooled 850 times faster than if radiative cooling
dominated.  (Radiative cooling certainly dominates under normal conditions
since individual bursts obey equation 3 [Liang \& Kargatis 1996; Liang
1997].)  Adiabatic cooling from the expansion of the material in the jet
is a promising mechanism to provide this cooling (Liang 1997) and might
allow the lag/luminosity relation to be a broken power law with a downturn
to low luminosities.  Indeed, Daigne \& Mochkovitch (1998, Eq. 30) show
that adiabatic cooling will dominate for $\Gamma$ values below some
threshold.

\section{VARIABILITY/LUMINOSITY RELATION}

The variability/luminosity relation (Fenimore \& Ramirez-Ruiz 2000) is
\begin{equation}
        L = 10^{52} (V/0.01)^{2.5 \pm 1},
\end{equation}
(Schaefer, Deng, \& Band 2001).  My explanation for this correlation
between $L$ and $V$ is that both quantities are strong functions of the
jet's
bulk relativistic expansion $\Gamma$ and hence are themselves correlated.

Burst light curves are presumed to be composed of emission from
disconnected regions each of which generates a sub-pulse that combine to
produce the pulses in the burst light curve (Fenimore {\it et
al.} 1999).  Throughout the time of the burst, the total number of
emitting
regions is $N_{total}$.  Within the internal shock scenario, the time at
which
BATSE sees the light from each region is dictated by events in the
collapsing object and hence does not depend on the jet's $\Gamma$.  So the
observed duration of the burst or peak pulse ($\Delta T$) will also be
independent
of $\Gamma$.  However, the observed duration of emission from each region
($\tau$) will scale as $\Gamma^{-2}$ (Fenimore, Madras, \& Nayakshin
1997).

The variability $V$ arises from fast fluctuations that will depend on the
average number of regions emitting at any given time ($N$).  The
$N_{total}$
sub-pulses of duration $\tau$ will be distributed over a time $\Delta T$,
so that 
\begin{equation}
	N = N_{total} (\tau / \Delta T).
\end{equation}
Since $\tau$ is proportional to $\Gamma^{-2}$ while the other factors do
not vary with $\Gamma$,
\begin{equation} 
	N \propto \Gamma^{-2}.
\end{equation}	
If there are few regions emitting at any instant, then the light curve
will be spiky because individual regions will appear relatively
isolated.  Alternatively, if $N$ is large, then the many independent
regions
will blur together to create a smooth light curve.  The rms scatter of the
light curve will scale as the square root of $N$.  Thus the variance of
the
light curve will vary as $N$.  The variability is the variance divided by
the square of the peak count rate (Fenimore \& Ramirez-Ruiz 2000) which
scales as the square of $N$, so V will scale as
\begin{equation}
	V \propto N^{-1}.
\end{equation}
This equation just quantifies the idea that if many independent regions
contribute at any given time then the light curve should be smooth.  From
equations 11 and 12,
\begin{equation}
	V \propto  \Gamma^2.
\end{equation}
As $\Gamma$ decreases, the emission from each region will broaden in time
and
merge to form a smooth light curve.

The luminosity of a burst will also depend on the jet's $\Gamma$.  The
total
energy available will be $\Gamma M_{jet}c^2$ for constant efficiency, and
this will be
seen by BATSE over a time of $\Delta T$.  For $\Theta_{jet} <
\Gamma^{-1}$, the entire front surface
of the jet will be visible from Earth so the luminosity should be
$\Gamma M_{jet}c^2/ \Delta T$.  However,  three correction factors are
needed.  The first
factor is $4 \Gamma^2$, which corrects for the relativistic beaming of the
radiation into a cone of opening angle $\Gamma^{-1}$.  The second factor
is
$(1+z)^{2-\alpha}$, which is a K-correction for the cosmological
expansion.  Here,
the average burst spectrum is taken as a power law of $E^{-\alpha}$, while
$\alpha$ is
typically around 2 (Schaefer et al. 1994, 1998).  So the second correction
factor is approximately unity on average.  The third factor is
$\Gamma^{\alpha}$, which
corrects for the red shift of the radiation in the BATSE energy band from
the relativistic motion of the jet towards Earth.  (There is no time
dilation correction for the relativistic motion of the jet towards us,
since the times of sub-pulses from individual regions are governed by
processes in the rest frame of the exploding source.  That is, each
sub-pulse will become narrower as $\Gamma$ increases, but the sub-pulses
are seen
by BATSE at the same times so the overall pulse made of smeared sub-pulses
will have the same peak luminosity.)  So for $\alpha \sim 2$, the third
factor is roughly $\Gamma^2$.  Taken together,
\begin{equation} 
	L = 4 \Gamma^2 \times 1 \times \Gamma^2
           \times (\Gamma M_{jet}c^2)/\Delta T.
\end{equation}
In all, the luminosity scales as
\begin{equation}
	L \propto  \Gamma^5.
\end{equation}
So as $\Gamma$ increases, the available energy and the fraction of that
energy
received by BATSE increases to a high power.

Both $L$ and $V$ depend on the $\Gamma$ value, and hence are correlated
through $\Gamma$.  From equations 13 and 15, 
\begin{equation}
	L \propto  V^{2.5}.
\end{equation}
This compares favorably with the observed variability/luminosity relation,
in that the exponent of equation
16 ($2.5$) is well within the uncertainties for the exponent of equation 9
($2.5 \pm 1$). 

In summary, I have simple and general explanations for the lag/luminosity
relation and the variability/luminosity relation.  When combined with the
successful prediction that a particular lag/variability relation should be
seen for 112 independent BATSE bursts, we can have every confidence in the
use of the two distance indicators.  So now we can get distances to GRBs
from gamma ray light curves alone (with no need for counterparts), and
this opens up a wide range of demographic studies involving all the many
BATSE bursts.


\begin{thebibliography}{}

\bibitem{daigne} Daigne, F. \& Mochkovitch, R. 1998, MNRAS, 296, 275.

\bibitem{fen99} Fenimore, E. E. 1999, ApJ, 518, 375.

\bibitem{fcrsyn99} Fenimore, E. E., Cooper, C., Ramirez-Ruiz, E., Sumner,
M. C., Yoshida, A., \& Namiki, M. 1999, ApJ, 512, 683.

\bibitem{claudine} Fenimore, E. E., Madras, C., \& Nayakshin, S. 1997,
ApJ, 473, 998.

\bibitem{fenimore} Fenimore, E. E. \& Ramirez-Ruiz, E. 2000, ApJ, submitted 
(astro-ph/0004176).

\bibitem{galama} Galama, T. J. et al. 1999, Nature, 398, 394.

\bibitem{kulkarni} Kulkarni, S. 2000, 20th Texas Symposium on Relativistic
Astrophysics, Austin.

\bibitem{liang} Liang, E. P. 1997, ApJ, 491, L15.

\bibitem{Karagatis} Liang, E. P. \& Kargatis, V. 1996, Nature, 381, 49.

\bibitem{norris} Norris, J. P., Marani, G., \& Bonnell, J. 2000, ApJ, 534,
248.

\bibitem{sdb} Schaefer, B. E., Deng, M., \& Band, D. L. 2001, ApJ,
submitted.

\bibitem{schaefer} Schaefer, B. E. et al. 1994, ApJSupp, 92, 285.

\bibitem{schaefer} Schaefer, B. E. et al. 1998, ApJ, 492, 696.

\bibitem{wsf} Walker, K. C., Schaefer, B. E., \& Fenimore, E. E. 2000,
ApJ, 537, 264.

\end{thebibliography}
\end{document}